\documentclass[prd,onecolumn,amsmath,amssymb,floatfix,nofootinbib]{revtex4}

\usepackage{amssymb}
\usepackage{amsmath}

\usepackage{sans}

\usepackage{graphicx}
\usepackage{graphics}
\usepackage{dcolumn}
\usepackage{color}
\usepackage{rotate}
\usepackage{fancyhdr}
\usepackage{hyperref}
\usepackage{indentfirst}

\usepackage{umoline}
\usepackage{ulem}

\def\beq{\begin{equation}}
\def\eeq{\end{equation}}
\def\be{\begin{eqnarray}}
\def\ee{\end{eqnarray}}
\def\ba{\begin{eqnarray}}
\def\ea{\end{eqnarray}}
\def\no{\nonumber}

\def\L{\mathcal{L}}

\def\M{\mathcal{M}}

\newcommand{\G}{\mathcal{G}}
\newcommand{\Z}{\mathcal{Z}}
\newcommand{\Y}{\mathcal{Y}}

\definecolor{darkred}{rgb}{.743,0,0}

\def\n02b{$0\nu\beta\beta$}
\def\n02bphi{$0\nu\beta\neta\phi$}
\def\lsim{\mathrel{\rlap{\lower4pt\hbox{\hskip1pt$\sim$}}
    \raise1pt\hbox{$<$}}}         %less than or approx. symbol
\def\gsim{\mathrel{\rlap{\lower4pt\hbox{\hskip1pt$\sim$}}
    \raise1pt\hbox{$>$}}}         %greater than or approx. symbol

\begin{document}
\title{Neutrinoless double-beta decay with massive scalar emission}

\author{Kfir Blum}
\affiliation{Weizmann Institute, Department of Particle Physics and Astrophysics, Rehovot, Israel 7610001}
\affiliation{CERN, Theoretical Physics Department, Switzerland}
\author{Yosef Nir}
\affiliation{Weizmann Institute, Department of Particle Physics and Astrophysics, Rehovot, Israel 7610001}
\author{Michal Shavit}
\affiliation{Weizmann Institute, Department of Particle Physics and Astrophysics, Rehovot, Israel 7610001}
%\date{\today}

\begin{abstract}
Searches for neutrino-less double-beta decay ($0\nu2\beta$) place an important constraint on models where light fields beyond the Standard Model participate in the neutrino mass mechanism. While $0\nu2\beta$ experimental collaborations often consider various massless majoron models, including various forms of majoron couplings and multi-majoron final-state processes, none of these searches considered the scenario where the ``majoron" $\phi$ is not massless, $m_\phi\sim$~MeV, of the same order as the $Q$-value of the $0\nu2\beta$ reaction. We consider this parameter region and estimate $0\nu2\beta\phi$ constraints for $m_\phi$ of order MeV. The constraints are affected not only by kinematical phase space suppression  but also by a change in the signal to background ratio charachterizing the search. As a result, $0\nu2\beta\phi$ constraints for $m_\phi>0$ diminish significantly below the reaction threshold. This has phenomenological implications, which we illustrate focusing on high-energy neutrino telescopes. The spectral shape of high-energy astrophysical neutrinos could exhibit features due to resonant $\nu\nu\to\phi\to\nu\nu$ scattering. Such features fall within the sensitivity range of  IceCube-like experiments, if $m_\phi$ is of order MeV, making $0\nu2\beta\phi$  a key complimentary laboratory constraint on the scenario. Our results motivate a dedicated analysis by $0\nu2\beta$ collaborations, analogous to the dedicated analyses targeting massless majoron models.
\end{abstract}
%\pacs{}

\maketitle
%\tableofcontents

%%%%%%%%%%%%%%%%%%%
\section{Introduction}
Neutrinoless double beta ($0\nu2\beta$) decay~\cite{Doi:1985dx,Rodejohann:2011mu,Vergados:2012xy,Pas:2015eia},
\beq
(A,Z)\to (A,Z+2)+2e^{-},
\eeq
is a lepton number violating process. It is sensitive to the neutrino mass parameter
\beq
m_{ee}=\left|\sum_i m_{\nu_i} U_{ei}^2\right|,
\eeq
where $m_{\nu_i}$ ($i=1,2,3$) are the neutrino masses and $U$ is the lepton mixing matrix~\cite{Patrignani:2016xqp}. While the renormalizable Standard Model (SM) has lepton number as an accidental symmetry and, consequently, predicts that the neutrinos are massless, adding dimension-five terms~\cite{Weinberg:1979sa}\footnote{Repeated flavour indices are summed-over, and the bracket $(HL)$ denotes contraction to an SU(2) singlet.}\footnote{See Ref.~\cite{Cirigliano:2017djv} for a recent discussion of $0\nu2\beta$ in the SM effective field theory.},
\beq\label{eq:dimfive}
{\cal L}_{d=5}=-\frac{\Z_{\alpha\beta}}{\Lambda}(HL_\alpha) (HL_\beta),
\eeq
where $H$ is the Higgs doublet field and $L_\alpha$ ($\alpha=e,\mu,\tau$) are the lepton doublet fields, leads to neutrino masses,
\beq
m_\nu = \frac{v^2\Z}{\Lambda},
\eeq
with $v=246$~GeV. 

We do not know the beyond-SM origin of the dimension-five terms in Eq.~(\ref{eq:dimfive}). It is possible that additional light particles accompany the neutrino mass mechanism and interact with SM fields in various ways. If there exists a light gauge-singlet scalar $\phi$, then the dimension-six terms 
\beq\label{eq:dimsix}
{\cal L}_{d=6}=-\frac{\Y_{\alpha\beta}}{\Lambda^2}\phi (HL_\alpha) (HL_\beta)
\eeq
are possible.
The dimension-six terms lead to Yukawa couplings of $\phi$ to neutrinos,
\be\label{eq:Lint}\L&\supset&-\frac{1}{2}\G_{\alpha\beta}\,\phi\,\nu_\alpha\nu_\beta+{\rm h.c.},\\
\label{eq:Gab}\G&=&\frac{v^2}{\Lambda^2}\Y.
\ee
If the mass of the $\phi$ particle is less than the $Q$-value of the $(A,Z)\to (A,Z+2)$ transition, $m_\phi<Q$, then the $\G_{ee}$ coupling leads to a decay where $0\nu2\beta$ is accompanied by on-shell $\phi$ emission  ($0\nu2\beta\phi$),
\beq
(A,Z)\to (A,Z+2)+2e^-+\phi.
\eeq

A well known framework that leads to Eq.~(\ref{eq:dimsix}) and to the decay mode $0\nu2\beta\phi$ is that of majoron models~\cite{PhysRevD.25.774,Georgi:1981pg,Gelmini:1980re,Vergados:1981gk,Chikashige:1980ui,Burgess:1992dt}, 
where $\phi$ is the Goldstone boson related to the spontaneous breaking of the lepton number symmetry. Many variants of the majoron model have been studied in the literature. In the simplest realisations, the seesaw scale $\Lambda$ appearing in Eq.~(\ref{eq:dimfive}) is promoted to a dynamical field, and the phase of this field is associated with $\phi$. In such models, (i) the $\phi$ particle is massless, and (ii) the terms of Eq.~(\ref{eq:dimfive}) and Eq.~(\ref{eq:dimsix}) are related, leading to $\G= m_\nu/\Lambda$. For high-scale seesaw models, with $\Z=\mathcal{O}(1)$, the seesaw scale is $\Lambda\sim10^{14}$~GeV, leading to $\G\sim10^{-24}$. As we review in Sec.~\ref{sec:0n2bphi}, such tiny coupling is some 20 orders of magnitude below the reach of $0\nu2\beta\phi$ searches.

In other scenarios, like the inverse-seesaw models of Ref.~\cite{PhysRevLett.56.561,PhysRevD.34.1642,BernabŽu1987303} (see Ref.~\cite{Weiland:2013wha} for a review), neutrino masses arise from effective dimension six terms. Namely, instead of $1/\Lambda$ in Eq.~(\ref{eq:dimfive}) we have $\mu/\Lambda^2$, where a technically natural hierarchy $\mu\ll\Lambda$ is responsible, at least in part, for the smallness of the neutrino mass. In such case, a light scalar field could arise if we promote the inverse-seesaw parameter $\mu$ to a field $\phi$ with $\mu=\langle\phi\rangle$. In this case, $\G= m_\nu/\mu$ and if $\mu$ is small enough, $0\nu2\beta\phi$ could be observable. If lepton number is broken spontaneously by $\langle\phi\rangle$, then the $\phi$ particle is still massless.

Global symmetries, however, are not expected to be exact. If lepton number is broken not only spontaneously 
but also explicitly, by some small parameter, then $\phi$ could be light but not massless~\cite{Rothstein:1992rh,Akhmedov:1992hi}. In addition, if the explicit lepton number violation (LNV) dominates the neutrino mass, then also the relation between $\G$ and $m_\nu$ is modified. Yet another framework that can accommodate this situation is if neutrinos are Dirac particles, in which case lepton number (more precisely some non-anomalous symmetry group containing it, e.g. $B-L$) may be exact; see~\cite{Berryman:2018ogk} for a recent study. 
While $0\nu2\beta$ experimental collaborations often consider various massless majoron models, such as different forms of the majoron-neutrino couplings and multi-majoron final-state processes, none of these searches considered the scenario of a massive majoron, $m_\phi\sim$~MeV, of the same order as the $Q$-value of the $0\nu2\beta$ reaction. In this paper we consider this parameter region\footnote{We note that Ref.~\cite{1993PhLB..308...85C} considered neutrino-less double-beta decay with emission of a massive vector boson.} and estimate $0\nu2\beta\phi$ constraints for the case of $m_\phi$ of order MeV. As we show, the constraints are affected not only by kinematical phase space suppression near $m_\phi\sim Q$, but also by a change in the signal to background ratio characterising the search. As a result, $0\nu2\beta\phi$ constraints for $m_\phi>0$ diminish significantly below the reaction threshold. Our results motivate a dedicated analysis by $0\nu2\beta$ collaborations, analogous to the dedicated analyses targeting different massless majoron models.

The constraint on massive $\phi$ emission in $0\nu2\beta\phi$ has phenomenological implications, which we illustrate focusing on high-energy neutrino telescopes.
Light scalar fields coupled to neutrinos were considered as mediators of anomalous neutrino self-interactions in many other works. Refs.~\cite{Goldberg:2005yw,Baker:2006gm} studied the effect of light scalar exchange on the energy spectrum of $\sim$10~MeV neutrinos from core-collapse supernovae (see also~\cite{Farzan:2014gza} where supernovae neutrinos scatter on dark matter).  Vector boson or massless majoron exchange were considered in~\cite{Kolb:1987qy,Keranen:1997gz,Hooper:2007jr}. 
Refs.~\cite{Chacko:2003dt,Hall:2004yg,Davoudiasl:2005ks,Friedland:2007vv} discussed the relation of anomalous neutrino interactions to low-scale neutrino mass generation, focusing on spontaneously broken global and gauged lepton number.
Ref.~\cite{Blum:2014ewa} extended the discussion to the technically natural possibility of small explicit LNV, and made a connection to phenomenology at high-energy neutrino telescopes.  
Recently, Ref.~\cite{Farzan:2018gtr} considered light scalar exchange in coherent neutrino-nucleus scattering.

Before we turn into concrete calculations, let us emphasize that while $0\nu2\beta$ is a LNV process, $0\nu2\beta\phi$ could be lepton number conserving (LNC). It could be therefore that the latter is strongly enhanced compared to the former. Explicitly, for $m_\phi\ll Q$, we have
\beq\label{eq:0n2bphi20n2bphi}
\frac{\Gamma_{0\nu2\beta\phi}}{\Gamma_{0\nu2\beta}}\sim\frac{|\G_{ee}|^2}{(4\pi)^2}\frac{Q^2}{m_{ee}^2}\gsim60\left(\frac{|\G_{ee}|}{10^{-5}}\right)^2,
\eeq
where we used the fact that $m_{ee}\lsim0.1$ eV~\cite{Barabash:2014uqa}, with $Q\sim$MeV. As we review in the next section, $0\nu2\beta\phi$ searches have reached a limit $|\G_{ee}|\lesssim10^{-5}$ (for massless $\phi$). The reason that $\Gamma_{0\nu\beta\beta\phi}\gg\Gamma_{0\nu\beta\beta}$, as shown by Eq.~(\ref{eq:0n2bphi20n2bphi}), is consistent with these limits, is related to the difference in the visible electron energy spectrum between these decay modes, which reduces the signal to background ratio for $0\nu2\beta\phi$ compared with $0\nu2\beta$. In what follows we will see an anaolgous effect deteriorating the sensitivity to $m_\phi>0$ compared to the $m_\phi=0$ case.

Finally, note that when $m_\phi>Q$, the on-shell process $0\nu2\beta\phi$ is kinematically blocked, but the off-shell process $0\nu2\beta(\phi^*\to2\nu)$, where a virtual $\phi$ is emitted and decays to two neutrinos, is always allowed. However, compared to the on-shell process $0\nu2\beta\phi$ (when allowed), the off-shell $\phi$ process is strongly suppressed by a factor $\sim\frac{2|\mathcal{G}|^2}{15(4\pi)^2}\,\frac{Q^4}{m_\phi^4}\lesssim10^{-7}\left(\frac{\mathcal{G}}{10^{-2}}\right)^2$.  In addition, the spectral shape with respect to the outgoing electron energy is  similar to that of the standard background process $2\nu2\beta$. These features are explained in App.~\ref{app:virt}. As a result of these features, the virtual $\phi$ decay mode $0\nu2\beta(\phi*\to2\nu)$ cannot be constrained with current experiments, and we limit our attention to on-shell $0\nu2\beta\phi$.

%%%%%%%%%%%%%%%%%%%
\section{Neutrinoless double-beta decay with massive scalar emission}\label{sec:0n2bphi}

From the list of $0\nu2\beta$ experiments surveyed in~\cite{Barabash:2014uqa}, NEMO-3~\cite{Arnold:2013dha} using $^{100}$Mo has the highest $Q$-value, $Q\approx3.03$~MeV. Recent work by NEMO-3 allowed them to surpass this record using $^{150}$Nd~\cite{Arnold:2016qyg}, with $Q\approx3.37$~MeV, albeit with lower exposure. Thus in principle $^{100}$Mo and $^{150}$Nd  experiments probe the highest scalar mass. The strongest constraint on the massless majoron case is from KamLAND-Zen~\cite{Gando:2012pj} using $^{136}$Xe, which has a somewhat lower value of $Q\approx2.5$~MeV. For these reasons -- sensitivity to the highest $m_\phi$, and current best sensitivity to massless $\phi$ -- we focus on $^{100}$Mo, $^{150}$Nd, and $^{136}$Xe in our numerical analysis below. It is straightforward to extend our analysis to other isotopes of common use, like $^{76}$Ge~\cite{GERDA:2018zzh,Aalseth:2017btx}, $^{82}$Se~\cite{Azzolini:2018dyb}, and $^{130}$Te~\cite{Andringa:2015tza,Alduino:2017ehq}. These isotopes yield comparable, although (currently) somewhat weaker constraints.

Refs.~\cite{Arnold:2013dha},~\cite{Arnold:2016qyg}, and~\cite{Gando:2012pj} provide 90\%CL bounds on the massless majoron scenario, equivalent to $|\G_{ee}|<(1.6-4.1)\times10^{-5}$, $|\G_{ee}|<(3.8-14.4)\times10^{-5}$, and $|\G_{ee}|<(0.4-1.0)\times10^{-5}$, respectively. 
These bounds are stronger than other constraints in the literature such as those arising from light meson decay (see, e.g.~\cite{Barger:1981vd,Lessa:2007up,Pasquini:2015fjv}) and from cosmological and astrophysical considerations~\cite{Blum:2014ewa}. 
While the massless majoron bounds~\cite{Doi:1987rx,Rath:2016saq,Hirsch:1995in,Pas:1999fc,Pas:2000vn} coincide with our model for $m_\phi\ll Q$, to our knowledge a study of the kinematical  region $m_\phi\sim Q\sim$~MeV has not been done and we consider this region in what follows. 

Searches for $0\nu2\beta\phi$ constrain the half-life time for the decay, $T_{\frac{1}{2}}$, which in our model can be approximately decomposed as~\cite{Doi:1985dx,Doi:1988zz,Doi:1987rx} (see, e.g.~\cite{Kotila:2015ata} for a recent account)
\be \label{eq:T12}T_{\frac{1}{2}}^{-1}&\approx&\left|\G_{ee}\right|^2\,\left|\M\right|^2\,G(m_\phi),\ee
where $\G_{ee}$ is defined in Eq.~(\ref{eq:Gab}), $\M$ is a dimensionless nuclear matrix element (NME), and $G(m_\phi)$ is a kinematical phase space factor, conventionally expressed in units of yr$^{-1}$.
The main effect of massive $\phi$ emission is to modify the phase space factor~\cite{Burgess:1992dt}, $G(0)\to G(m_\phi)$:
\be\label{eq:G} G(m_\phi)&\propto&
\int dE_1\int dE_2 \,E_1p_1\,E_2p_2\,\left(\left(Q+2m_e-E_1-E_2\right)^2-m_\phi^2\right)^{\frac{1}{2}}\,a\left(E_1,E_2\right)
.\ee
The outgoing electrons Coulomb factor, encoded in $a(E_1,E_2)$, is given in Refs.~\cite{Doi:1985dx,Doi:1987rx}. 
For the range of electron momenta of interest, $p_i>0.5$ MeV, $a(E_1,E_2)$ can be factorized as $a(E_1,E_2)\approx F_0(E_1)F_0(E_2)$. For simplicity, for most of the numerical results in this work we use the non-relativistic approximation,
\be
\label{eq:PRA}F(v)&\approx&\frac{2\pi\alpha Z_f/v}{1-e^{-2\pi\alpha Z_f/v}}.
\ee
We have checked that our results are not affected significantly when using the relativistic expressions for the electron wave function.

\begin{figure}[htbp]
\begin{center}
\includegraphics[width=0.8\textwidth]{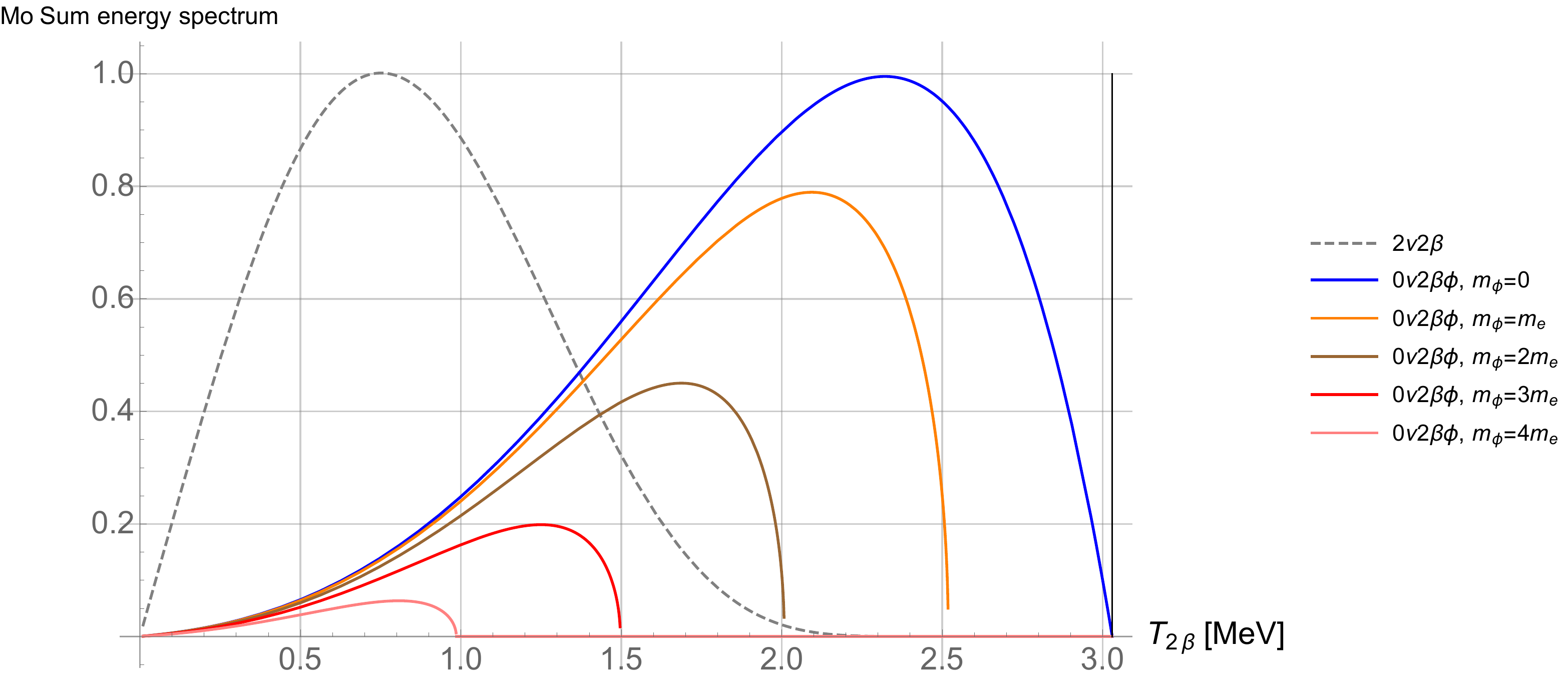}
\caption{Energy spectrum (sum of kinetic energies of the two electrons) for $0\nu2\beta\phi$. The grey dashed curve (solid vertical line) shows the spectrum for $2\nu2\beta$ ($0\nu2\beta$) decay. The normalisation is arbitrary, apart from the relative normalisation of the $0\nu2\beta\phi$ curves which follow the phase space factor.}
\label{fig:spectrum}
\end{center}
\end{figure}

The $0\nu2\beta\phi$ constraint on $\G_{ee}$ deteriorates as the scalar mass $m_\phi$ approaches the kinematical limit for the decay. The constraints are affected in two ways:
\begin{enumerate}
\item The phase space factor diminishes close to the kinematical limit, $G(m_\phi)/G(0)\to0$ as $m_\phi\to Q$.
\item As $m_\phi$ is increased, the visible final state electrons kinetic energy
\be T_{2\beta}=E_1+E_1-2m_e\ee
is pushed to lower values, because the usual massless majoron phase space factor $(Q-T_{2\beta})$ is replaced by $((Q-T_{2\beta})^2-m_\phi^2)^{\frac{1}{2}}$. This brings the distribution of $T_{2\beta}$ to overlap more with the $T_{2\beta}$ distribution of the irreducible Standard Model $2\nu2\beta$ background, leading to smaller signal to background ratio in the experimentally relevant energy range.
\end{enumerate}
Fig.~\ref{fig:spectrum} illustrates both of the above points, showing the visible electron $T_{2\beta}$ spectrum for $0\nu2\beta\phi$ for different values of $m_\phi$ and comparing to the $2\nu2\beta$ case (in arbitrary normalisation). This plot considers $^{100}$Mo as an example.
Note that the experimental analyses typically impose a lower energy cutoff of $T_{2\beta}\gtrsim0.5$~MeV.
%
%%%%%%%%%%%%%%%%%%%
%%%%%%%%%%%%%%%%%%%

\begin{figure}[htbp]
\begin{center}
\includegraphics[width=0.55\textwidth]{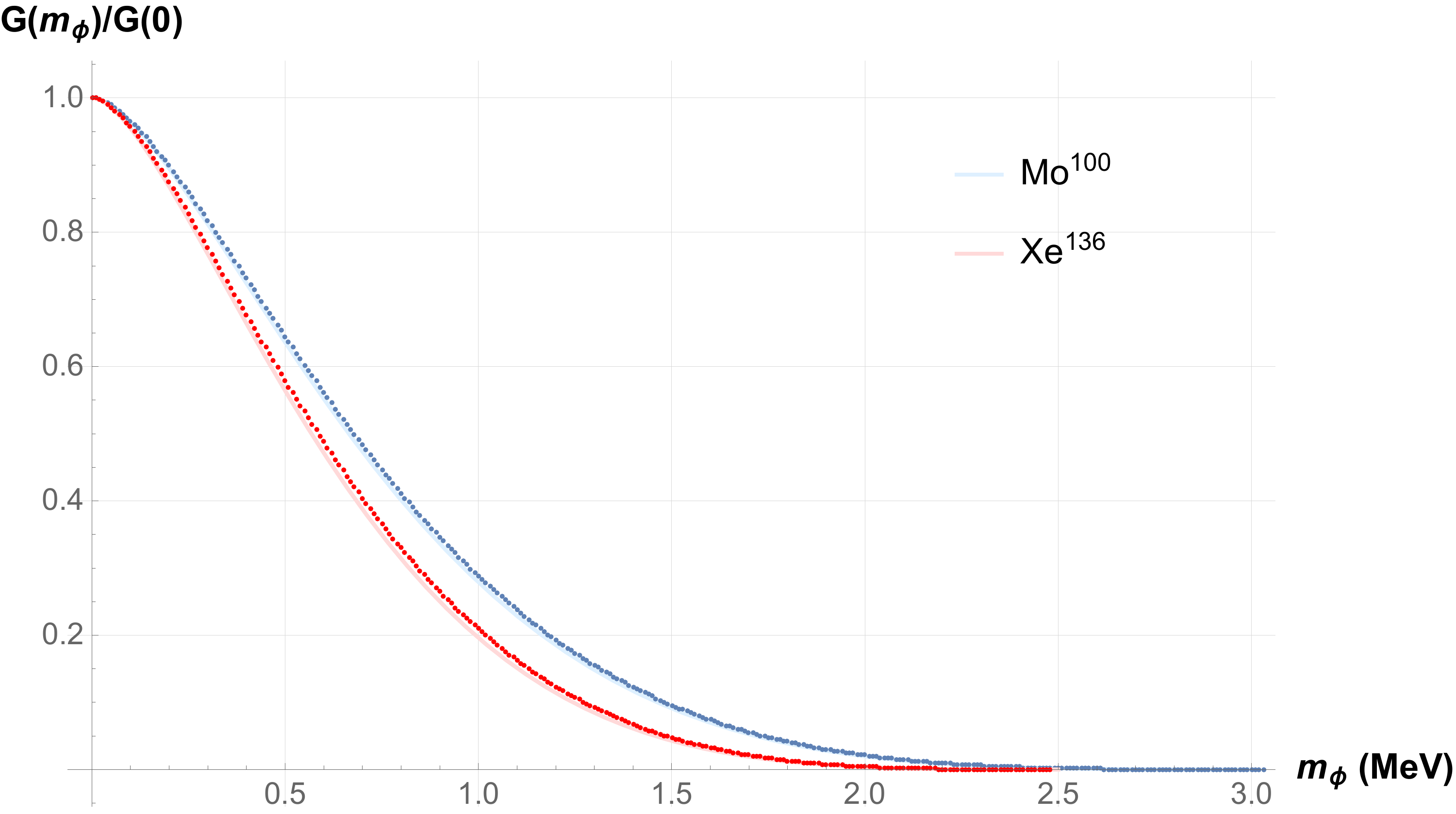}
\caption{Phase space suppression factor for massive $\phi$, vs. $m_\phi$, for $^{100}$Mo (blue) and $^{136}$Xe (red). The analytical PRA result Eq.~(\ref{eq:Gm2G0PRA}) is shown in solid lines. Numerical computation using the full relativistic electron wave function is shown by dots.}
\label{fig:Gm2G0}
\end{center}
\end{figure}

Proceeding further, we consider first the simple phase space suppression encoded in $G(m_\phi)/G(0)$. In the Primakoff-Rosen approximation (PRA)~\cite{0034-4885-22-1-305}, we can calculate this ratio analytically. We find:
\be\label{eq:Gm2G0PRA}
\frac{G\left(m_{\phi}\right)}{G(0)}&=&\frac{1}{40}\frac{1}{210\bar{m}_{e}^{4}+210\bar{m}_{e}^{3}+84\bar{m}_{e}^{2}+14\bar{m}_{e}+1}\nonumber\\
&\times&\bigg[40\sqrt{1-\bar{m}_{\phi}^{2}}\left(210\bar{m}_{e}^{4}+210\bar{m}_{e}^{3}+84\bar{m}_{e}^{2}+14\bar{m}_{e}+1\right)\\
 &+&2\sqrt{1-\bar{m}_{\phi}^{2}}\left(8400\bar{m}_{e}^{4}+27300\bar{m}_{e}^{3}+23240\bar{m}_{e}^{2}+6790\bar{m}_{e}+759\right)
 \bar{m}_{\phi}^{2}\nonumber \\
 & +&\sqrt{1-\bar{m}_{\phi}^{2}}\left(\left(8960\bar{m}_{e}^{2}+7910\bar{m}_{e}+1779\right)\bar{m}_{\phi}^{4}
 +128\bar{m}_{\phi}^{6}\right)\nonumber \\
 & -&\log\left(\frac{1+\sqrt{1-\bar{m}_{\phi}^{2}}}{\bar{m}_{\phi}}\right)840\bar{m}_{\phi}^{2}\left(30\bar{m}_{e}^{4}
 +60\bar{m}_{e}^{3}+40\bar{m}_{e}^{2}+10\bar{m}_{e}+1\right)\nonumber \\
 & -&\log\left(\frac{1+\sqrt{1-\bar{m}_{\phi}^{2}}}{\bar{m}_{\phi}}\right)2100\left(6\bar{m}_{e}^{3}+12\bar{m}_{e}^{2}
 +6\bar{m}_{e}+1\right)\bar{m}_{\phi}^{4}+105\left(10\bar{m}_{e}+5\right)\bar{m}_{\phi}^{6}\bigg],\nonumber
\ee
where $\bar{m}_{\phi}= m_{\phi}/Q$ and $\bar{m}_{e}= m_{e}/Q$.
Taking the limit $\bar{m}_{\phi}\rightarrow0$,
we recover the phase space factor for massless majoron (see, e.g.~\cite{Georgi:1981pg}). The phase space suppression factor is shown in Fig.~\ref{fig:Gm2G0} for $^{100}$Mo (blue) and $^{136}$Xe (red). The analytical PRA result Eq.~(\ref{eq:Gm2G0PRA}) is shown in solid lines. Numerical computation using the full relativistic electron wave function is shown by dots.

We see that phase space suppression is appreciable already for $m_\phi\sim1$~MeV. In the case of $^{136}$Xe ($^{100}$Mo), for $m_\phi\approx2$~MeV (2.5 MeV), the decay width drops to $\sim$1\% of its value for the massless majoron case. This means that the limit on $\G_{ee}$ becomes weaker than the massless majoron limit by a factor of at least $\sim$10 at that point, even ignoring the signal to background ratio deterioration effect that we discuss later on. 
%
%%%%%%%%%%%%%%%%%%%
%%%%%%%%%%%%%%%%%%%

We next consider the varying signal to background ratio at varying $m_\phi$.
Accounting for this effect properly is difficult outside of the experimental collaboration, due, among other factors, to different experiment-dependent sources of background, which make the derived limit sensitive to the signal spectral shape. Our discussion here provides a crude approximation of the limits, and motivates a dedicated analysis by the experimental collaborations, analogous to the analysis done when setting limits on various massless majoron models with different spectral indices.

\begin{figure}[htbp]
\begin{center}
\includegraphics[width=0.75\textwidth]{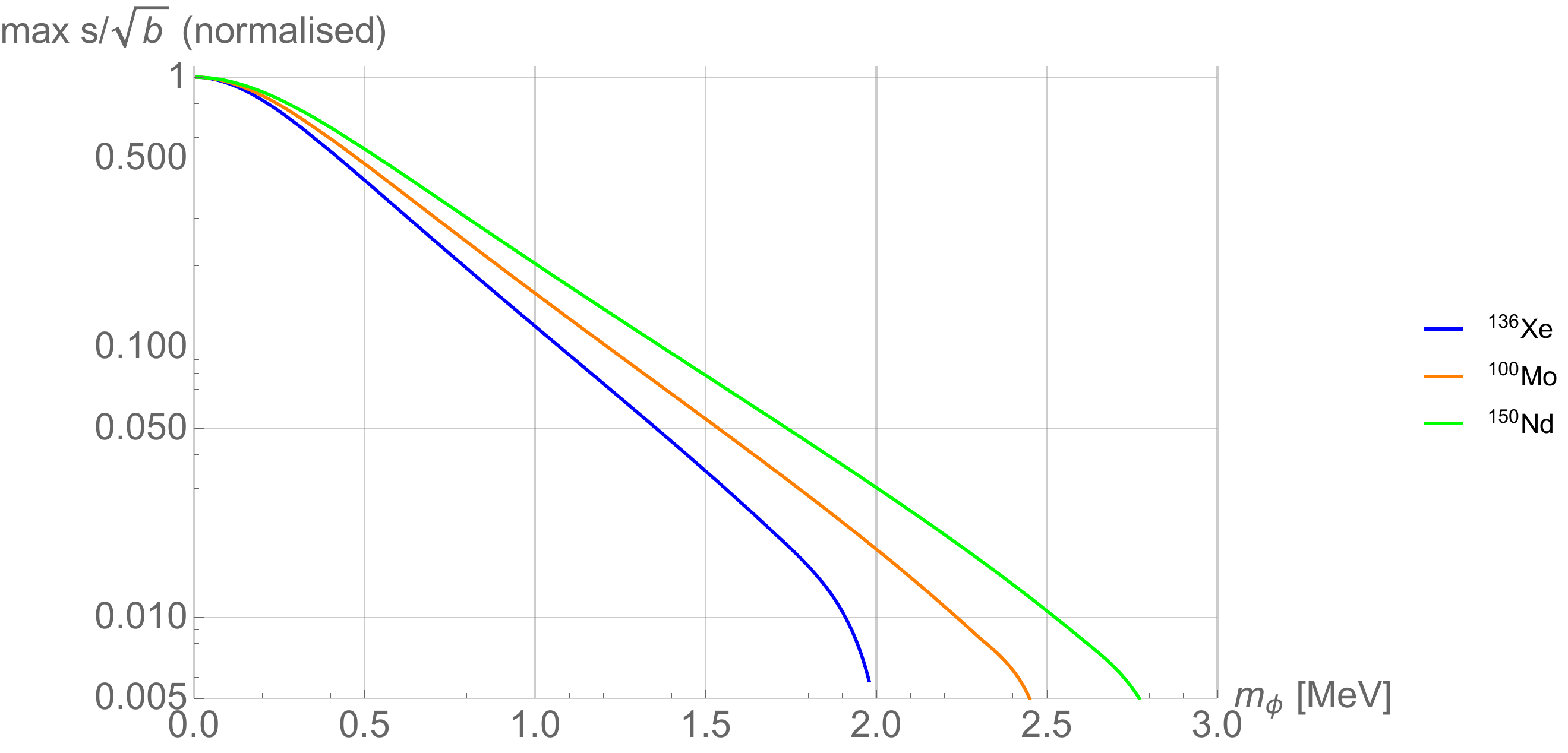}
\caption{Normalised signal to root-background, as function of $m_\phi$.}
\label{fig:s2rb}
\end{center}
\end{figure}

We can estimate the spectral effect on the limits by considering $s/\sqrt{b}$, where for background ($b$) we take the $2\nu2\beta$ spectrum and for signal ($s$) the spectrum of $0\nu2\beta\phi$ with the selected value of $m_\phi$. The limit would deteriorate approximately in proportion to the maximum value of $s/\sqrt{b}$, attained in the region $T_{2\beta}\gtrsim0.5$~MeV analyzed by the experiments.
This information is presented in Fig.~\ref{fig:s2rb}. The x-axis of the plot gives the scalar mass $m_\phi$. The y-axis gives the value of $s/\sqrt{b}$, where we calculate this ratio at the point\footnote{In this calculation we do not account for the energy resolution of KamLAND-Zen and of NEMO-3~\cite{Gando:2012pj,Arnold:2013dha}. We checked that this does not modify the results appreciably.} in $T_{2\beta}$ that maximizes $s$ within the range $T_{2\beta}>0.5$~MeV. 
To make the interpretation easier, we normalize max $s/\sqrt{b}$ obtained at any value of $m_\phi$ to the value of max $s/\sqrt{b}$ obtained for $m_\phi=0$. Fig.~\ref{fig:s2rb} then implies that the $^{136}$Xe limit on $T_{\frac{1}{2}}$ for $m_\phi=1$~MeV, for example, should be about a factor of 10 weaker than the limit on $T_{\frac{1}{2}}$ for $m_\phi=0$.
%%%%%%%%%%%%%%%%%%%
%%%%%%%%%%%%%%%%%%%

We stress that the exercise in Fig.~\ref{fig:s2rb} is a rough approximation only: the official experimental analyses contain additional important sources of background from various radioactive contaminants, that are typically fit alongside with the signal. Nevertheless, in what follows we set estimated limits in the parameter space of $\mathcal{G}_{ee}$ and $m_\phi$, using the max $s/\sqrt{b}$ information from Fig.~\ref{fig:s2rb} together with the phase space suppression factor $G(m_\phi)/G(0)$. The lower bound we take for $T_{\frac{1}{2}}$ is 
\be T^{\rm limit}_{\frac{1}{2}}(m_\phi)&=&\frac{{\rm max}\left\{s/\sqrt{b}(m_\phi)\right\}}{{\rm max}\left\{s/\sqrt{b}(0)\right\}}\,T^{\rm limit}_{\frac{1}{2}}(0),\ee
where $T^{\rm limit}_{\frac{1}{2}}(0)$ is the limit placed by the collaboration for the massless $\phi$ case. This translates to an upper bound on $\mathcal{G}_{ee}$ that reads (here and elsewhere, where experimental limits are considered they always refer to the absolute value $|\G_{ee}|$)
\be\label{eq:cstrtGee}\mathcal{G}_{ee}^{\rm limit}(m_\phi)&=&\sqrt{\frac{G(0)}{G(m_\phi)}\,\frac{{\rm max}\left\{s/\sqrt{b}(0)\right\}}{{\rm max}\left\{s/\sqrt{b}(m_\phi)\right\}}}\,\mathcal{G}_{ee}^{\rm limit}(0).\ee

\begin{figure}[htbp]
\begin{center}
\includegraphics[width=0.75\textwidth]{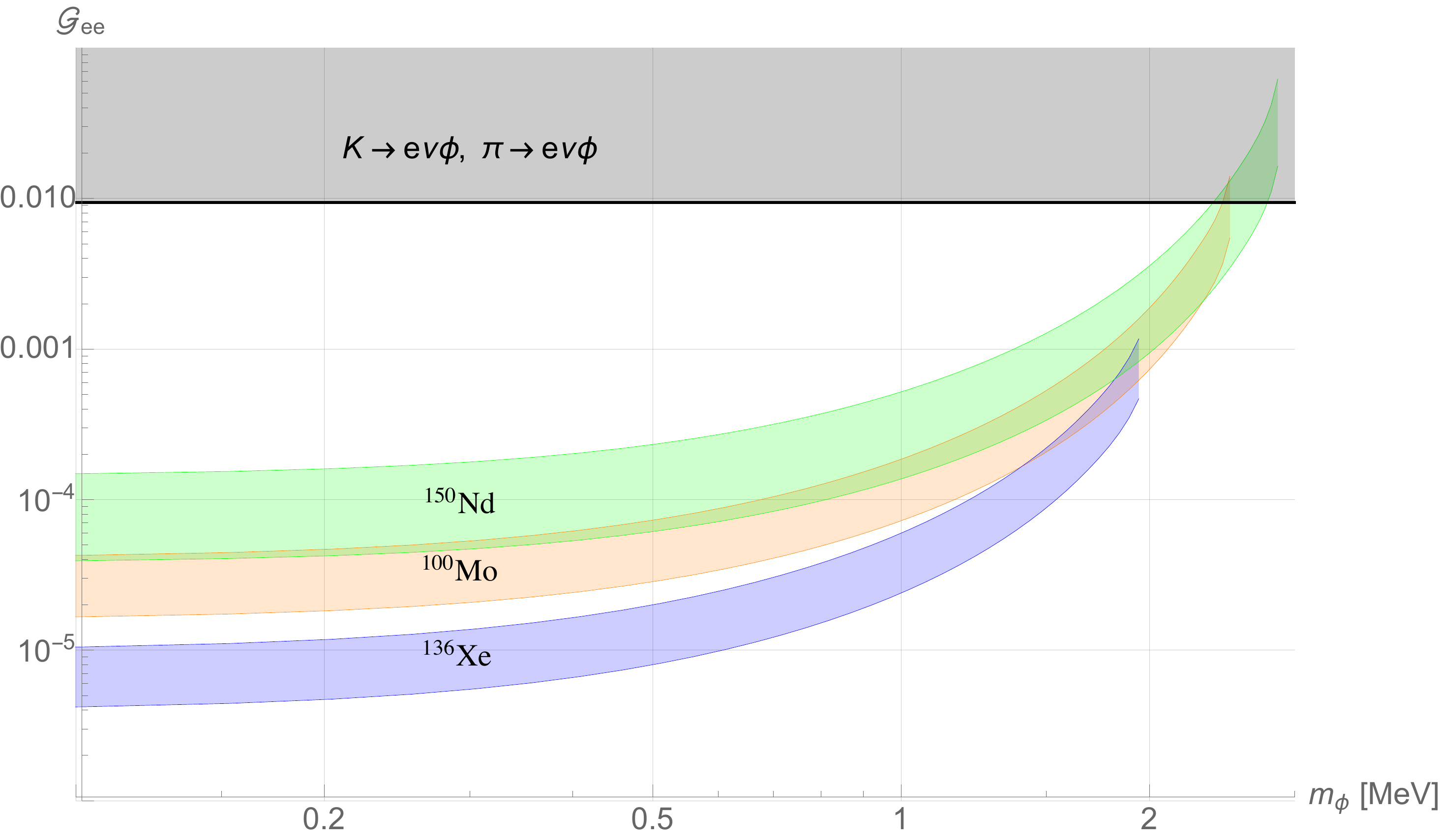}
\caption{90\%CL upper bound on $\G_{ee}$. The region above the coloured shaded bands is excluded by KamLAND-Zen~\cite{Gando:2012pj} (blue, $^{136}$Xe), NEMO-3~\cite{Arnold:2013dha} (orange, $^{100}$Mo), and NEMO-3~\cite{Arnold:2016qyg} (green, $^{150}$Nd). The width of the band represents the uncertainties quoted by the collaborations regarding the massless $\phi$ limits; we stress that our approximate signal to background analysis implies larger uncertainty at $m_\phi>0$. For comparison with other constraints, the dark shaded region above the horizontal black line shows the constraint from light meson decays~\cite{Pasquini:2015fjv}.}
\label{fig:GeeLim}
\end{center}
\end{figure}

In Fig.~\ref{fig:GeeLim} we plot the 90\%CL upper bound on $\G_{ee}$, evaluated using Eq.~(\ref{eq:cstrtGee}). The region above the shaded bands is excluded by KamLAND-Zen~\cite{Gando:2012pj} (blue, $^{136}$Xe), NEMO-3~\cite{Arnold:2013dha} (orange, $^{100}$Mo), and NEMO-3~\cite{Arnold:2016qyg} (green, $^{150}$Nd). The width of the band represents the uncertainties quoted by the collaborations for the massless $\phi$ limits. We stress that our approximate signal to background analysis implies larger uncertainty at $m_\phi>0$. For comparison with other constraints, the dark shaded region above the horizontal black line shows the constraint from light meson decays~\cite{Pasquini:2015fjv}.
%
%%%%%%%%%%%%%%%%%%%
%%%%%%%%%%%%%%%%%%%

\section{Implications for scalar-mediated neutrino self-interactions}
Neutrino-neutrino interactions through light mediator exchange can cause observable features in the diffuse high-energy neutrino flux seen by IceCube and future neutrino telescopes~\cite{Ng:2014pca,Ioka:2014kca,Blum:2014ewa}. This can occur if resonant s-channel scattering of a high-energy astrophysical neutrino with energy $\epsilon_\nu$ off the cosmic neutrino background (C$\nu$B) is possible, which in turn requires the mediator mass to match the center of mass energy (CME) of the collision,
\be m_\phi&=&\sqrt{2m_\nu\epsilon_\nu}=2\left(\frac{\epsilon_\nu}{100~\rm TeV}\right)^{\frac{1}{2}}\left(\frac{m_\nu}{0.02~\rm eV}\right)^{\frac{1}{2}}~{\rm MeV},\ee
where $m_\nu$ is the mass of the C$\nu$B neutrino participating in the collision.
High-energy neutrino telescopes like IceCube and its planned Gen2 upgrade would be most sensitive to features in the astrophysical neutrino flux in the energy range from a few tens of TeV (below which the atmospheric background kicks in) up to around PeV (above which statistics fall off).
The MeV mediator mass range is therefore of particular interest to this phenomenology.

The detailed connection between $0\nu2\beta\phi$ limits and high-energy neutrino phenomenology is model-dependent. For concreteness, in the rest of this section we consider the framework of Ref.~\cite{Blum:2014ewa}, where the coupling $\G_{\alpha\beta}$ was proportional to the neutrino mass matrix. In particular, each real non-negative neutrino mass eigenvalue $m_{\nu_i}$ is accompanied by a real non-negative value of $\G_i=\G_{\alpha\beta}U_{\alpha i}U_{\beta i}$, while off-diagonal terms vanish in the mass basis, $\G_{\alpha\beta}U_{\alpha i}U_{\beta j}=0$ for $i\neq j$.
The optical depth for resonant scattering, $\nu_i\nu_i\to\phi\to\nu\nu$, proceeding through the scalar $\Phi$ in Eq.~(\ref{eq:Lint}), considering neutrinos with observed energy $\epsilon_\nu$, is bounded approximately by
\footnote{Eq.~(\ref{eq:optd}) is an upper bound to $\tau_{\nu\nu\to\phi\to\nu\nu}$ because it assumes that the scalar $\phi$ can only decay back to neutrinos via Eq.~(\ref{eq:Lint}), minimizing its width; if other decay modes are possible for $\phi$, the resonant scattering cross section and the optical depth are suppressed by $\Gamma_{\phi\to\nu\nu}/\Gamma_\phi$.}
\be\label{eq:optd}\tau_{\nu_i\nu_i\to\phi\to\nu\nu}&\lesssim&2\,\left(\frac{\mathcal{G}_i}{10^{-4}}\right)^2\left(\frac{m_\phi}{2~\rm MeV}\right)^{-2}\left(\frac{m_\phi^2}{2m_{\nu_i}\epsilon_\nu}\right)^3\theta\left(\frac{m_\phi^2}{2m_{\nu_i}}-\epsilon_\nu\right)\theta\left((1+z)\epsilon_\nu-\frac{m_\phi^2}{2m_{\nu_i}}\right).%\nonumber\\
\ee
In Eq.~(\ref{eq:optd}), the high-energy neutrino is assumed to have been emitted at redshift $z$. The relevant astrophysical emission is typically thought to be dominated around $z\sim2$ or so (see, e.g.~\cite{Waxman:1998yy,Bahcall:1999yr,Waxman:2013zda,Halzen:2013dva,Meszaros:2014tta}). Detectable effects at neutrino telescopes require $\tau=\mathcal{O}(1)$, implying $\G_i\gsim10^{-4}$.

\begin{figure}[htbp]
\begin{center}
\includegraphics[width=0.6\textwidth]{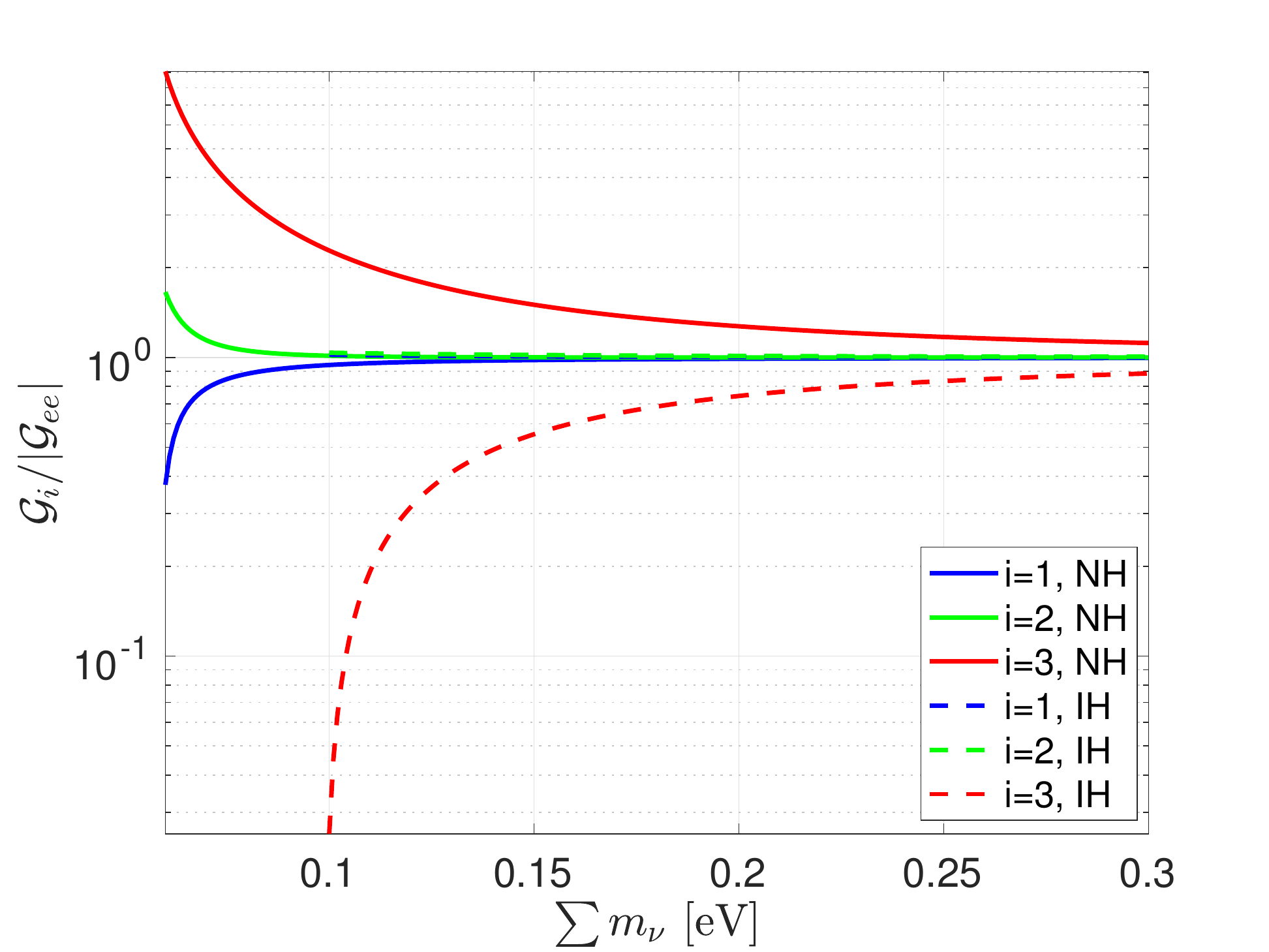}
\caption{The ratio $\G_i/|\G_{ee}|$ for the different neutrino mass eigenstates $i=1,2,3$, for normal and inverted hierarchy, as function of the sum of neutrino mass $\sum m_\nu$.}
\label{fig:Gi2Gee}
\end{center}
\end{figure}

In this model, because $\G_i\propto m_{\nu_i}$, we have
\be \G_i=|\G_{ee}|\,\left|\frac{m_{\nu_i}}{m_{ee}}\right|.\ee
If we are given the neutrino mass hierarchy (e.g. by upcoming neutrino oscillation experiments~\cite{Capozzi:2017ipn}) and the sum of neutrino masses (e.g. by cosmology~\cite{Lesgourgues:2006nd,Abazajian:2016yjj}), then, using the measured PMNS values, we can relate the bound on $\G_{ee}$ to a bound on $\G_i$ for any $i$. In Fig.~\ref{fig:Gi2Gee} we do this exercise, using neutrino oscillation parameters from Ref.~\cite{Patrignani:2016xqp}.
%%%%%%%%%%%%%%%%%%%
%%%%%%%%%%%%%%%%%%%

\begin{figure}[htbp]
\begin{center}
\includegraphics[width=0.75\textwidth]{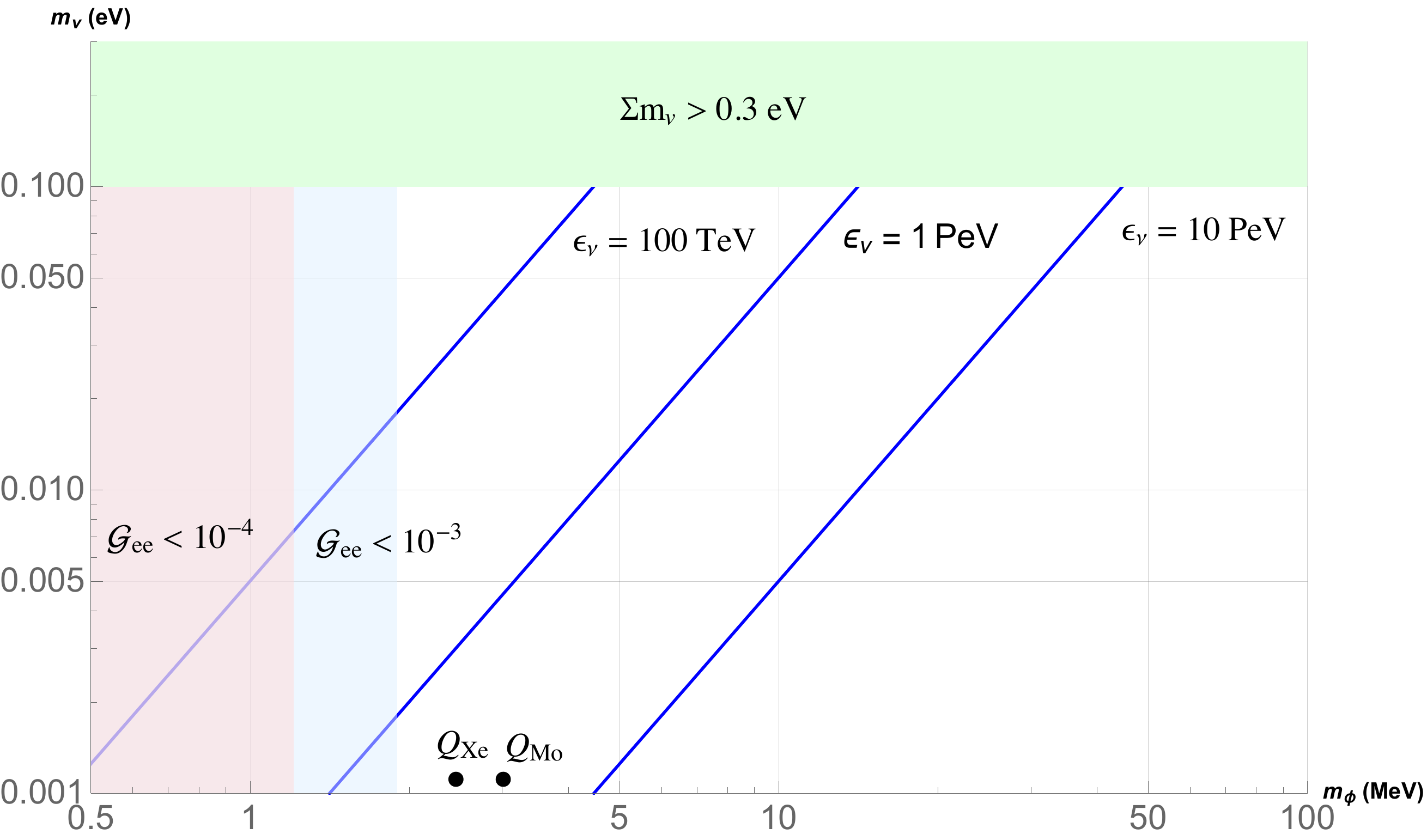}
\caption{$0\nu2\beta\phi$ constraints presented in the $m_\phi-m_\nu$ parameter space, relevant for high-energy neutrino phenomenology. In and to the left of the red (blue) shaded area, $\G_{ee}<10^{-4}$ ($\G_{ee}<10^{-3}$) at 90\%CL. Diagonal lines show the observer frame neutrino energy which enters resonance s-channel scattering for scalar exchange with a C$\nu$B neutrino of mass $m_\nu$. In and above the green shaded region, the sum of neutrino masses exceeds 0.3~eV and is excluded by cosmological observations. The $Q$-values for $0\nu2\beta$ in $^{100}$Mo and $^{136}$Xe are indicated by black dots at the bottom of the plot.}
\label{fig:mnu}
\end{center}
\end{figure}

Finally, in Fig.~\ref{fig:mnu} we present our result for the $0\nu2\beta\phi$ in the parameter space relevant for high-energy neutrino telescopes. In and to the left of the red (blue) shaded area, $\G_{ee}<10^{-4}$ ($\G_{ee}<10^{-3}$) at 90\%CL, where we have used the upper side (more conservative) of the KamLAND-Zen limit, rescaled from the $m_\phi=0$ case as shown in Fig.~\ref{fig:GeeLim}. As explained above, the constraint on $\G_{ee}$ can be readily converted to a constraint on neutrino optical depth, given information about the neutrino mass hierarchy and total mass. Diagonal lines show the observer frame neutrino energy which enters resonance s-channel scattering for scalar exchange with a C$\nu$B neutrino of mass $m_\nu$. In and above the green shaded region, the sum of neutrino masses exceeds 0.3~eV and is excluded by cosmological observations. The $Q$-values for $0\nu2\beta$ in $^{100}$Mo and $^{136}$Xe are indicated by black dots at the bottom of the plot. We add these indicators to signify the effect of the phase space and signal to background considerations, which cause the naive $0\nu2\beta\phi$ constraint for a massless $\phi$, $|\G_{ee}|\lesssim10^{-5}$, to deteriorate in the massive $\phi$ case.
%%%%%%%%%%%%%%%%%%%
%%%%%%%%%%%%%%%%%%%

\section{Conclusions}
A gauge-singlet scalar $\phi$ is expected to couple to two neutrinos, $\nu_\alpha\nu_\beta$ ($\alpha,\beta=e,\mu,\tau$) with couplings $\G_{\alpha\beta}$ suppressed by $v^2/\Lambda^2$, where $\Lambda$ is a scale of additional new physics. The $\G_{ee}$ coupling can lead to neutrino-less double beta decay accompanied by scalar emission, $0\nu2\beta\phi$. Experimental searches for $0\nu2\beta\phi$ have been conducted under the assumption that $\phi$ is the majoron, that is the massless Goldstone boson related to the spontaneous breaking of lepton number symmetry. It could, however, be the case that lepton number is explicitly broken, and $\phi$ is a massive scalar. The $0\nu2\beta\phi$ decay will proceed if $\G_{ee}\neq0$ and the decay is kinematically allowed, $m_\phi<Q$.

If $m_\phi$ is not much smaller than $Q$, then the bound on $\G_{ee}$ extracted from the experimental upper bound on $T_{\frac{1}{2}}^{-1}$, the $0\nu2\beta\phi$ decay rate, is weakened compared to the massless majoron case. In this work, we obtained these bounds by considering the two main relevant effects:
\begin{itemize}
\item The phase space factor $G(m_\phi)$ (see Fig. \ref{fig:Gm2G0}), which is suppressed compared to $G(0)$;
\item The reduction in $s/\sqrt{b}$ (see Fig. \ref{fig:s2rb}), the signal to root-background ratio, which is a consequence of the modification of the $T_{2\beta}$ spectrum ($T_{2\beta}$ is the sum of the kinetic energies of the two electrons).
\end{itemize}
The bounds on $\G_{ee}$ for the massive scalar case are presented in Fig. \ref{fig:GeeLim}.

The modification of the bounds from the massless majoron case to the massive scalar case are relevant for $m_\phi={\cal O}({\rm MeV})$. A scalar in this mass range which couples to neutrinos can have a strong effect on high energy astrophysical neutrinos observed by IceCube, as it mediates resonant scattering of these neutrinos on the cosmic neutrino background. Thus, $0\nu2\beta\phi$ constraints on massive scalars exclude part of the parameter space where the relevant features in the neutrino spectrum measured by IceCube may appear. This relation between $0\nu2\beta\phi$ and high energy neutrino phenomenology is presented in Fig. \ref{fig:mnu}.

The exciting possibility to discover gauge-singlet scalars with mass in the MeV range via $0\nu2\beta$ experiments (and the possible relation with high energy astrophysical neutrino observations) call for dedicated analyses by the experiments, where the effects of $m_\phi\neq0$ are carefully taken into account.

\acknowledgments
We thank Laura Baudis and Jacobo Lopez-Pavon for useful correspondence.
KB is incumbent of the Dewey David Stone and Harry Levine career development chair, and is supported by grant 1937/12 from the I-CORE program of the Planning and Budgeting Committee and the Israel Science Foundation and by grant 1507/16 from the Israel Science Foundation.
YN is the Amos de-Shalit chair of theoretical physics, and is supported by grants from the Israel Science Foundation (grant number 394/16), the United States-Israel Binational Science Foundation (BSF), Jerusalem, Israel (grant number 2014230), the I-CORE program of the Planning and Budgeting Committee and the Israel Science Foundation (grant number 1937/12), and the Minerva Foundation.

\begin{appendix}
\section{Phase space for different decay modes}\label{app:virt}
In the main text, it was convenient to decompose the inverse decay lifetime for the $0\nu2\beta\phi$ process as (reproducing Eq.~(\ref{eq:T12}))
\be \label{eq:T12mtxt}T_{\frac{1}{2}}^{-1}&\approx&\left|\G_{ee}\right|^2\,\left|\M\right|^2\,G(m_\phi),\ee
where $G(m_\phi)$ is the kinematical factor, encoding the important features of the $0\nu2\beta\phi$ mode, and $\mathcal{M}$ is the nuclear matrix element.

Here we reproduce this decomposition in some more detail, in order to compare the different processes $0\nu2\beta$, $0\nu2\beta\phi$, and the off-shell scalar process $0\nu2\beta(\phi*\to2\nu)$. In the latter process, we refer to the case where no neutrinos are emitted from the weak current terms in the direct nuclear decay calculation (as in $0\nu2\beta\phi$), but where the scalar mass $m_\phi$ is larger than the reaction threshold, leading to off-shell scalar diagram that can be decomposed as $0\nu2\beta\phi^*$ with virtual $\phi$ decaying via $\phi^*\to\nu\nu$.

For $0\nu2\beta$, we have
\be\Gamma_{\tilde M\to Mee}&\approx&\frac{1}{2M}|\mathcal{M}_{\tilde M\to Mee}|^2\int\frac{d^3P}{(2\pi)^32M}\frac{d^3k_1}{(2\pi)^32\epsilon_1}\frac{d^3k_2}{(2\pi)^32\epsilon_2}\,k_1\cdot k_2\,(2\pi)^4\delta^{(4)}\left(P+k_1+k_2-\tilde P\right)\,a(\epsilon_1,\epsilon_2)\no\\
&=&\frac{(4\pi)^2}{(2\pi)^516M^2}|\mathcal{M}_{\tilde M\to Mee}|^2\int d\epsilon_1k_1d\epsilon_2k_2\,\epsilon_1\epsilon_2\,\delta\left(Q+2m_e-\epsilon_1-\epsilon_2\right)\,a(\epsilon_1,\epsilon_2).\ee

For $0\nu2\beta\phi$, with on-shell $\phi$, we have
%%%%
\be\Gamma_{\tilde M\to Mee\phi}&\approx&\frac{1}{2M}|\mathcal{M}_{\tilde M\to Mee\phi}|^2\int\frac{d^3P}{(2\pi)^32M}\frac{d^3k_1}{(2\pi)^32\epsilon_1}\frac{d^3k_2}{(2\pi)^32\epsilon_2}\frac{d^3k}{(2\pi)^32\epsilon}\,k_1\cdot k_2\,(2\pi)^4\delta^{(4)}\left(P+k_1+k_2+k-\tilde P\right)\,a(\epsilon_1,\epsilon_2)\no\\
&=&\frac{(4\pi)^3}{(2\pi)^832M^2}|\mathcal{M}_{\tilde M\to Mee\phi}|^2\int d\epsilon_1k_1d\epsilon_2k_2\,\epsilon_1\epsilon_2\, \left((Q+2m_e-\epsilon_1-\epsilon_2)^2-m_\phi^2\right)^{\frac{1}{2}}\,a(\epsilon_1,\epsilon_2),\ee
in which $G(m_\phi)$ of Eq.~(\ref{eq:G}) can be identified.

Finally, for $0\nu2\beta(\phi*\to2\nu)$ we have
%%%%
\be\label{eq:virt}\Gamma_{\tilde M\to Mee(\phi^*\to\nu\nu)}&\approx&\frac{1}{2M}|\mathcal{M}_{\tilde M\to Mee\phi}|^2\int\frac{d^3P}{(2\pi)^32M}\frac{d^3k_1}{(2\pi)^32\epsilon_1}\frac{d^3k_2}{(2\pi)^32\epsilon_2}\frac{d^3k_{\nu_1}}{(2\pi)^32\epsilon_{\nu_1}}\frac{d^3k_{\nu_2}}{(2\pi)^32\epsilon_{\nu_2}}\,k_1\cdot k_2\no\\
&\times&\frac{|\mathcal{M}_{\phi\to\nu\nu}|^2}{((k_{\nu_1}+k_{\nu_2})^2-m_\phi^2)^2+m_\phi^2\Gamma_\phi^2}\,(2\pi)^4\delta^{(4)}\left(P+k_1+k_2+k_{\nu_1}+k_{\nu_2}-\tilde P\right)\,a(\epsilon_1,\epsilon_2)\no\\
&=&
\frac{(4\pi)^3}{(2\pi)^{8}32M^2}|\mathcal{M}_{\tilde M\to Mee\phi}|^2\int d\epsilon_1k_1\,d\epsilon_2k_2\,\epsilon_1\epsilon_2\,a(\epsilon_1,\epsilon_2)\,\int_0^\infty d\epsilon\,\delta\left(Q+2m_e-\epsilon_1-\epsilon_2-\epsilon\right)\no\\
&\times&\frac{\epsilon^5}{m_\phi^5}\,\frac{2\Gamma_\phi}{\pi}\int_0^1 dx\frac{x^2\,(1-x^2)}{\left(1-\frac{\epsilon^2}{m_\phi^2}(1-x^2)\right)^2+\frac{\Gamma_\phi^2}{m_\phi^2}}\no\\
&\approx&
\frac{(4\pi)^3}{(2\pi)^{8}32M^2}|\mathcal{M}_{\tilde M\to Mee\phi}|^2\int d\epsilon_1k_1\,d\epsilon_2k_2\,\epsilon_1\epsilon_2\,a(\epsilon_1,\epsilon_2)\,\frac{4\Gamma_\phi}{15\pi}\,\frac{\left(Q+2m_e-\epsilon_1-\epsilon_2\right)^5}{m_\phi^5}\left(1+\mathcal{O}\left(\frac{Q^2}{m_\phi^2}\right)\right).\no\\
\ee
%
%%%%
Here, the decay width of $\phi$ into two neutrinos is given by
\be\Gamma_\phi&=&\frac{|\mathcal{G}|^2}{2m_\phi}\int\frac{d^3k_{\nu_1}}{(2\pi)^32\epsilon_{\nu_1}}\frac{d^3k_{\nu_2}}{(2\pi)^32\epsilon_{\nu_2}}\,k_{\nu_1}\cdot k_{\nu_2}\,(2\pi)^4\delta^{(4)}(k_{\nu_1}+k_{\nu_2}-k_\phi)=
\frac{|\mathcal{G}|^2m_\phi}{32\pi},\ee
with $|\mathcal{G}|^2=\sum_i\mathcal{G}_i^2$.

Considering the virtual $\phi$ process, Eq.~(\ref{eq:virt}), we see that:
\begin{itemize}
\item Compared to the on-shell process $0\nu2\beta\phi$, the decay rate for the off-shell process $0\nu2\beta(\phi*\to2\nu)$ is suppressed by a factor $\sim\frac{2|\mathcal{G}|^2}{15(4\pi)^2}\,\frac{Q^4}{m_\phi^4}\sim10^{-7}\left(\frac{\mathcal{G}}{10^{-2}}\right)^2\frac{Q^4}{m_\phi^4}$. This suppression can be recognised as the product of (i) an additional final state phase space factor, (ii) an insertion of $\mathcal{G}^2$, and (iii) an over-all kinematical factor $\sim Q^4/m_\phi^4$.
\item Besides from the over-all suppression of the $0\nu2\beta(\phi*\to2\nu)$ process, the spectral shape with respect to the outgoing electron energy is of the form $\sim \left(Q+2m_e-\epsilon_1-\epsilon_2\right)^5$, which is, of course, just the spectral shape of the standard background process $2\nu2\beta$.
\end{itemize}
As a result of these features, the virtual $\phi$ decay mode $0\nu2\beta(\phi*\to2\nu)$ cannot be constrained with current experiments.

\end{appendix}

\vspace{6 pt}

\bibliography{ref}

\end{document}